\documentclass[sigconf]{acmart}
\usepackage[justification=centering]{caption}
\AtBeginDocument{%
  \providecommand\BibTeX{{%
    \normalfont B\kern-0.5em{\scshape i\kern-0.25em b}\kern-0.8em\TeX}}}

\renewcommand\footnotetextcopyrightpermission[1]{} 
\setcopyright{none}

\begin{document}

\title{Meltdown: Bridging the Perception Gap in Sustainable Food Behaviors Through Immersive VR}

\author{Acacia Chong Xiao Xuan}
\email{e0773988@u.nus.edu}
\affiliation{%
    \institution{National University of Singapore}
    \country{Singapore}
}

\author{Florentiana Yuwono}
\email{florentianayuwono@u.nus.edu}
\affiliation{%
    \institution{National University of Singapore}
    \country{Singapore}
}

\author{Melissa Anastasia Harijanto}
\email{melissaharijanto@u.nus.edu}
\affiliation{%
    \institution{National University of Singapore}
    \country{Singapore}
}

\author{Xu Yi}
\email{xu.yi@u.nus.edu}
\affiliation{%
    \institution{National University of Singapore}
    \country{Singapore}
}



\begin{abstract}
Climate change education often struggles to bridge the perception gap between everyday actions and their long-term environmental consequences. In response, we developed Meltdown, an immersive virtual reality (VR) escape room that simulates a grocery shopping and food waste management experience to educate university students in Singapore about sustainable consumption. The game emphasizes sustainable food choices and disposal practices, combining interactive elements and narrative feedback to promote behavioral change. Through a user study with 36 university students, we observed statistically significant improvements in participants’ objective knowledge, perceived confidence, and intention to adopt sustainable behaviors. Our results suggest that experiential VR environments can enhance climate education by making abstract environmental concepts more immediate and personally relevant.
\end{abstract}




\maketitle

\section{Introduction}
Climate change is a pressing global crisis and individual consumer choices contribute significantly to its acceleration \cite{MaciejKolaczkowski2021}. From food selection to packaging preferences, everyday decisions have profound environmental impacts. In Singapore, a densely populated city state dependent on imports, practicing sustainable consumption is crucial to mitigating carbon footprints and resource depletion. Recent reports highlight the growing concern about food waste and unsustainable packaging in the region \cite{nea_2023}\cite{tan_2020}. To address these issues, we are targeting university students, who are at a critical stage of forming long-term consumer habits. By targeting this group, our aim is to test sustainable consumption behaviors on a smaller scale before expanding to the wider public. 

Educating people about climate change remains a challenge, largely due to the perception gap. The long-term effects of climate change often feel abstract and distant, which makes it difficult for individuals to connect their current actions with future outcomes. This disconnect can reduce the sense of urgency and personal relevance, limiting the effectiveness of traditional educational approaches.

\subsection{Current Solutions}
According to our research, the current solutions are the following: 
\begin{itemize}
    \item \textbf{Information campaigns and labeling}: Many organizations, including the Singapore Environment Council, provide information on eco-friendly products and labeling schemes such as the Singapore Green Label \cite{mse_2024}. However, information alone may not translate into behavioral change. 
    \item \textbf{Mobile Apps and Checklists}: Apps like "SusGain" \cite{scholarship_guide_2021} offer rewards and information on sustainable choices. Although effective for some, they may not provide an immersive or engaging experience that fosters long-term behavioral changes and does not provide understanding on why something is considered sustainable.

    There are also more applications that provide a gamified experience with sustainable checklists, such as GoBeEco \cite{novo_zanchetta_goldmann_vaz_2024}, but are limited to a 2D experience.
    \item \textbf{Educational Workshops and Seminars}: Workshops and seminars on sustainable living, often conducted by community centers and NGOs, provide in-depth knowledge but may not reach a wide audience, as they are primarily held by individual communities such as the South East Community Development Council \cite{moe_2025}.
    \item \textbf{Retailer Initiatives}: Supermarkets like FairPrice in Singapore have introduced initiatives such as the promotion of reusable bags and the reduction of plastic packaging \cite{fairprice_group_2023}. These efforts are important, but require active participation from consumers.
\end{itemize}
\subsection{Our Solution}
Our solution is a fully immersive virtual reality supermarket game that simulates a real-life grocery shopping experience. Users are tasked with making sustainable food choices, considering aspects such as packaging and transportation. We chose to focus particularly on sustainable food choices, as our preliminary survey (detailed in Sections 4 and 5) revealed that this is the area in which participants felt the least knowledgeable and perceived the greatest need for education. To deepen the learning experience, we also included a post-shopping kitchen segment where users sort their food waste and learn sustainable disposal methods, such as composting fruit and vegetable peels in a vinegar jar, drying coffee grounds for reuse as cleaning agents, grinding eggshells into protein powder, and properly discarding bones. 

By experiencing both the selection and disposal processes in a realistic virtual environment, users develop a more comprehensive understanding of the environmental impact of their daily choices. This immersive experience attempts to \textbf{reduce the perception gap}, helping users connect their current actions with long-term environmental consequences, making the issues more immediate and tangible. This approach is more effective than traditional methods because it provides an experiential and engaging learning environment, encouraging continuous behavioral change. Intuitively, the immersive nature of VR allows users to "live" the consequences of their choices, leading to more impactful learning than passive information consumption.

Our primary research hypothesis is that \textbf{VR experiences can immerse university students living in Singapore in scenarios to educate them about climate change and motivate them to take steps to reduce their contribution to climate change.}

We aim to investigate whether this experience can translate into tangible changes in real-world behavior within the Singapore context, before potentially expanding the application to a global scale. This approach aims to address the limitations of current 2D informational or gamified solutions, particularly in the crucial area of sustainable food choices.

\subsection{Contributions}
\begin{enumerate}
    \item We developed an \textbf{immersive VR escape room experience} that simulates sustainable grocery shopping and food waste disposal, specifically designed to address climate change education among university students in Singapore.
    \item Through a user study with 36 participants, we found that the VR experience led to \textbf{statistically significant improvements in sustainability-related knowledge}, confidence, and behavioral intentions.
    \item Our findings demonstrate \textbf{the potential of VR as an effective medium for bridging the perception gap} in climate education and promoting long-term behavioral change.
\end{enumerate}

\subsection{Organization of the Next Sections}
\begin{itemize}
    \item \textbf{Section 2: Related Works} - This section reviews existing research and technologies related to the use of VR supermarkets, as well as existing climate change VR games.
    \item \textbf{Section 3: Game Design and Implementation} - This section details the design and development of the VR supermarket game, including the virtual environment, interaction mechanics, and sustainability metrics.
    \item \textbf{Section 4: Evaluation Method} - This section describes the user study design, data collection methods, and evaluation metrics used to assess the effectiveness of the game.
    \item \textbf{Section 5: Evaluation Results} - This section presents the findings of the user study, including quantitative and qualitative data on user behavior and learning outcomes.
    \item \textbf{Section 6: Acknowledgments} - This section expresses gratitude to individuals and organizations that contributed to the research and development of the VR supermarket game, but are not listed as authors.
    \item \textbf{Section 7: Conclusion} - This section summarizes the key contributions of the work and reiterates the potential of VR to promote continuous sustainable behavior.
\end{itemize}

\section{Related Works} 

\subsection{Using VR Supermarket for Nutritional Research and Education: A Scoping Review \cite{cristiana_2025}} 
    Virtual reality supermarket simulations have gained traction within nutritional research, due to challenges associated with conducting controlled studies in real-world supermarkets \cite{cristiana_2025}. The key difference lies in the primary educational objective. The application in the paper focuses on promoting nutritional awareness and healthier food selections, while our work focuses on educating on the environmental consequences of food waste and non-sustainable packaging, and its impact on climate change.

\subsection{Cleanopolis VR \cite{tan_nurul‐asna_2023}}
Cleanopolis VR is a VR game where players learn about CO2 with a fictional character to improve air quality in a generalized world. While both use VR for environmental education, our application differs by focusing on a realistic, everyday activity – grocery shopping. Instead of a fictional setting, users in our VR supermarket directly experience the environmental impact of their real-life choices concerning food, packaging, and transportation. This grounding in tangible, real-world scenarios distinguishes our approach from Cleanopolis VR's more abstract and fictional context.

\subsection{Data-driven VR Applications to Visualize the Effects of Climate Change \cite{9815983} \cite{Huang03062021}}
Many VR climate change applications visualize data-driven impacts on natural environments such as rising sea levels \cite{9815983} and forests \cite{Huang03062021}. While valuable for research and awareness, their ability to foster direct behavioral change in daily life may be limited. This can stem from a lack of connection to actionable behaviors and the portrayal of environments less relevant to everyday experiences, potentially reducing personal engagement and perceived urgency.

\subsection{Design of 3D Virtual Reality in the Metaverse for Environmental Conservation Education Based on Cognitive Theory \cite{lo_tsai_2022}}
VRAM is a mobile VR-based educational platform that aims to improve learning about water and soil conservation through interactive 3D multimedia content. It allows users to explore educational materials by watching video clips, viewing annotations, and answering questions, creating an engaging, yet organized learning experience. Although VRAM is effective in helping users engage with content, our application takes it a step further by adding decision-making elements and showing real-time consequences. Instead of following a fixed sequence, users in our system actively shape their virtual environment through their choices, encouraging behavioral learning through direct feedback.

\subsection{Learning real-life cognitive abilities in a novel 360°-virtual reality supermarket: a neuropsychological study of healthy participants and patients with epilepsy \cite{grewe_2013}}
Recent VR supermarket studies, such as OctaVis, use immersive environments for neuropsychological training, improving cognitive performance through multi-layered learning \cite{grewe_2013}. These applications focus on cognitive rehabilitation, with immersion linked to improved task outcomes.

In contrast, our VR application targets climate change education by promoting sustainable shopping decisions and showing the real-world effects of unsustainable choices. Although existing VR supermarkets focus on cognitive skills, our approach uses VR to raise environmental awareness and directly engage users in sustainability. This shifts the focus from cognitive training to climate education.

\subsection{Distinction from Existing Work}
Our work extends existing VR supermarket applications by focusing on climate change education that links everyday behaviors and their environmental impact, providing real-time feedback on sustainable choices. This approach uniquely combines immersive learning with actionable behavior.

\section{VR System Design \&\ Implementation} 
Our solution, built using Unity 6.40.f1, is available on our GitHub repository \cite{Yuwono_Chong_Harijanto_Xu}.

\subsection{Main Features}

\subsubsection{\textbf{Player Interactions}}
Users can teleport, grab, and interact with certain objects by pressing the VR controller's trigger button.

\subsubsection{\textbf{Scenes}}
\begin{itemize}
    \item \textbf{Tutorial Room}: This is where users can learn to familiarize themselves with VR controls.
    \item \textbf{Supermarket}: Users navigate a virtual supermarket where they are tasked with making sustainable choices for grocery items based on factors such as packaging, origin, and transportation. The system tracks the environmental impact of each decision. Poor choices trigger a transition to the Disaster Room.
    \item \textbf{Transition Scene}: A transition scene between the Supermarket and Food Waste Room is added to provide context for the change in environment and task. It helps users mentally prepare for the next stage, reinforcing the idea that sustainability continues beyond the act of purchasing.
    \item \textbf{Food Waste Room/Kitchen}: After shopping, users enter a kitchen where they sort their food waste. Here, they interact with different waste items and learn how to dispose of each sustainably. Poor choices will also trigger a transition to the Disaster Room.
    \item \textbf{Disaster Room}: Users are presented with the consequences of their unsustainable choices through a combination of visual text overlays and narrated voiceovers. This immersive feedback helps users understand the environmental impact of their decisions and encourages reflection, allowing them to learn from their mistakes in a meaningful and engaging way.
    \item \textbf{End Scene}: Users can view their game completion status, including the time taken and scores achieved for the mechanisms in both the supermarket and kitchen sections.
\end{itemize}

\subsubsection{\textbf{GameObjects and Other Assets}}
\begin{itemize}
    \item \textbf{Food Prefabs}: Two similar game objects to indicate a more sustainable and a less sustainable choice
    \item \textbf{Scanner Prefab}: Users can interact and find out more about the food prefabs by scanning them. To do so, the user can grab the scanner and press the trigger button on either controller while pointing to the food prefab. 
    \item \textbf{Poster Prefab}: Users can interact with the poster by hovering on it and clicking the trigger button using the trigger button on either controller.
    \item \textbf{Panels and Overlays}: Show users the educational information about sustainability. These will appear when users interact with the object by scanning it, when they interact with posters, and when the users make the right choice. These will also appear in the disaster room to educate users why the choice they are making is not sustainable.
    \item \textbf{Background Music and Sound Effects}: To make users more immersed in the experience.
\end{itemize}


\subsection{Preventing VR Sickness}
\subsubsection{Smooth Locomotion Management}

This game primarily uses smooth locomotion for precise movement within a scene, which can sometimes induce motion sickness due to sensory conflicts. To counter this, a vignette effect has been applied during movement. The vignette effect darkens peripheral vision, reducing visual input to the rods of the eye, which are highly sensitive to motion in peripheral vision. By limiting excessive motion perception in the periphery, the vestibulo-ocular reflex is better aligned with the visual experience, reducing discomfort. Meanwhile, forward vision remains clear, which allows the user to rely more on the cones, providing better visual detail and colour perception. Since cones are concentrated in the fovea, they help to maintain a stable image without triggering undesired motion sensitivity. In addition, linear velocity is maintained, avoiding sudden acceleration or deceleration, which can cause vestibular discomfort.

\subsubsection{Teleportation Between Scenes}
Scene changes are handled through teleportation with a blackout effect to avoid abrupt transitions that could cause disorientation. This eliminates any unintended motion during scene loading, ensuring a smooth transition without disrupting the user's sense of balance.

\subsubsection{Interaction Design and Perceived Affordances}
In our VR environment, user interactions are guided by environmental affordances and naturalistic depth cues to create an intuitive and immersive experience. Objects and spaces are designed with clear cues, such as a door indicating a point of entry and exit from a room. Depth perception is enhanced through occlusion, shading, and realistic size variations, helping users judge distances and navigate the environment easily. Visual and audio feedback, like animations or sounds when interacting with objects, reinforces affordances, ensuring a seamless and engaging user experience without intrusive UI elements. Objects are also manipulated using natural grab-and-pull gestures to promote intuitive interaction and prevent visual-vestibular mismatches. Lastly, the user's hands are stylised in a robotic form to prevent the uncanny valley effect, ensuring smooth user acceptance.

\subsection{Usability Goals}
\begin{enumerate}
    \item \textbf{Effectiveness}: Users should be able to complete tasks such as selecting items and navigating the scene efficiently without confusion.
    \item \textbf{Efficiency}: The interaction flow is streamlined, reducing unnecessary movements and ensuring quick decision-making.
    \item \textbf{Safety}: The experience minimises sudden motions, visual clutter, and VR sickness triggers to ensure comfort for prolonged use.
    \item \textbf{Learnability}: New users are introduced gradually with progressive onboarding, ensuring they can quickly adapt to the VR interface.
    \item \textbf{Memorability}: The interface and interactions are designed to be consistent and intuitive, reducing the need for relearning.
    \item \textbf{Utility}: The game offers valuable insights into sustainable actions while maintaining an engaging, game-like experience.
    
\end{enumerate}

\subsection{Enhancements Based on User Feedback}

The following enhancements were made to improve clarity, interaction, and overall user experience:

\begin{itemize}
    \item \textbf{Repositioned eggs in the supermarket scene}
    
    The eggs are moved further away from the tomato prefab to prevent overlap scanning areas, allowing the user to scan them more easily.
    
    \item \textbf{Added audio narration when users make the correct choice in the supermarket}

    An audio narration is now added because many users overlooked the visual overlay explaining their correct choice. 

    \item \textbf{Adjusted player's starting position in the supermarket}
    
    The starting point was moved closer to the entrance to help guide users to the scanner, which is essential to compare the sustainability of different food options.
    
    \item \textbf{Highlighted key text on composting posters in the kitchen}
    
    Important keywords were highlighted on instructional posters to clarify that only one composting step is needed to progress.

    \item \textbf{Added subtitle for door transition cue}
    
    A brief on-screen subtitle reading "The door is open" now appears (and disappears after two seconds) to guide users to the kitchen after finishing the supermarket stage.
    
\end{itemize}

\subsection{Iterations}
\begin{enumerate}
    \item \textbf{Alpha Stage}

    In this stage, we conducted preliminary user research to understand user needs and preferences before the development phase. The development phase, which focused on creating scenes, mechanics, and game logic, took three weeks to complete.
    \item \textbf{Beta Stage}
    
    During the Beta stage, we performed internal developer tests and performed testing with a small group of users. This allowed us to gather feedback and refine key features based on user input, ensuring that the application was improving in line with user expectations.
    \item \textbf{Gold Stage}
    
    The Gold stage focuses on external play testing to collect feedback and data to refine the escape room. The objective is to identify usability issues, validate educational impact and improve the mechanics of the game, with the goal of fine-tuning the product for a polished and engaging experience.
    
\end{enumerate}

\section{Evaluation Method} 
\subsection{Preliminary User Research}
To understand the public (especially university students in Singapore) awareness of climate change initiatives in Singapore and to evaluate how knowledgeable they are about existing policies, we conducted a survey. The survey is aimed at gauging people's recognition of climate change prevention measures and includes questions on participants' familiarity with local climate change policies (e.g., the Singapore Green Plan 2030 and the National Climate Change Strategy, etc.), their current efforts to combat climate change, and their knowledge of large-scale initiatives such as carbon pricing and renewable energy investments. 

The results from this survey will help identify which policies are the most under-recognized and the gap in climate change education. This enables us to tailor our virtual reality application to address these knowledge gaps more effectively. The complete survey questions can be found in \textbf{Appendix ~\ref{appendix:preliminary_user_research_qns}}.

\subsection{Methods Used to Evaluate the Application}

\subsubsection{Product Use Case and Evaluation Methodology}
Our VR application is an educational escape room experience designed for university students in Singapore, with the aim of raising awareness and encouraging sustainable food-related behaviors. Users are involved in a simulated environment where they make real-time decisions about grocery shopping and food waste disposal.

We conducted a user study with \textbf{36 university students} from various academic backgrounds to evaluate both the usability and the educational effectiveness of the experience. Each session involved: 
\begin{itemize} 
    \item A short background interview to assess familiarity with VR. 
    \item A pre-play survey to gauge baseline knowledge and habits. 
    \item The full VR experience, including tutorial, supermarket, kitchen, and final segments. 
    \item A post-play survey and guided interview to capture user feedback. 
\end{itemize}

During the game, participants were encouraged to think aloud while facilitators observed interactions. The key usability aspects evaluated include: 
\begin{itemize} 
    \item \textbf{Intuitiveness of interaction}: How easily users navigated and interacted with objects. 
    \item \textbf{Clarity of instructions and feedback}: Whether the users understood the tasks in each room. 
    \item \textbf{Immersion and engagement}: How compelling and educational the experience was. 
    \item \textbf{Comfort and accessibility}: Including issues like VR motion sickness or navigation difficulty. 
\end{itemize}

Observational notes, structured questions, and participant ratings formed the basis of usability feedback. For the complete user testing protocol, refer to \textbf{Appendix ~\ref{appendix:user_testing_guide}}.

\subsubsection{Objective and Subjective Metrics}

We evaluated the VR application using our pre- and post- survey responses. These metrics allow us to assess both the effectiveness of educational content and the usability of the VR experience.

To comprehensively evaluate the impact of the VR experience, we categorized our findings into three domains:
\begin{enumerate}
    \item \textbf{Objective knowledge gains}: Measurable improvements in factual understanding of sustainability topics.
    \item \textbf{Subjective shifts in perceptions}: Changes in participants' current attitudes and confidence about sustainability.
    \item \textbf{Changes in intended behaviors}: Future-oriented intentions regarding sustainability actions.
\end{enumerate}

\textbf{Objective knowledge gains} were drawn from the knowledge-based survey questions given before and after the game.
\begin{itemize}
     \item Understanding key contributors to the environmental impact of food production.
     \item Ability to identify features of sustainable food packaging.
     \item Awareness of why reducing food waste is environmentally important.
\end{itemize}

To assess changes in participants' sustainability knowledge, we use the Wilcoxon signed-rank test (one-sided) to evaluate whether the VR experience led to an increase in correct responses, as the data is binary and paired.

Unlike traditional knowledge assessments where a normal distribution might be assumed, the binary nature and potential ceiling effects of our scoring method warranted a non-parametric approach. Furthermore, we adopted a significance threshold of \textbf{90\% (p < 0.10)} rather than the more conventional 95\%, based on these key factors:
\begin{enumerate}
    \item the relatively small sample size of 36 participants, which limits statistical power,
    \item the exploratory nature of our study, which aims to identify promising educational outcomes rather than make definitive population-wide claims, and
    \item the potential impact of prior knowledge, as some participants may already have baseline knowledge of climate change.
\end{enumerate}
Using a more lenient threshold allows us to detect meaningful trends that might be missed under stricter criteria.

Our null hypothesis is as follows:
\begin{quote}
\textbf{H\textsubscript{0}: There is no significant improvement in participants’ climate change knowledge after playing the VR game.}
\end{quote}

\textbf{Subjective shifts in perceptions} were collected using the Likert scale survey questions given before and after the game:
\begin{itemize}
    \item Perceived improvement in familiarity with the concept of food sustainability.
    \item Change in confidence when making environmentally friendly food choices while shopping.
    \item Shift in likelihood of considering sustainability during food-related purchasing decisions.
\end{itemize}

To measure shifts in perceptions and intentions, we used Likert scale items that asked participants to rate their confidence, familiarity, and commitment to sustainable behaviors before and after the VR experience. The Wilcoxon signed-rank test (two-sided) was applied to this ordinal, paired data to determine whether the observed changes were statistically significant.

We used a conventional \textbf{95\% significance level (p < 0.05)} here, as subjective self-report measures—such as confidence and intention—tend to be more variable and context-dependent. Maintaining a stricter threshold helps avoid over-interpreting potentially short-term or biased self-perceptions. Additionally, this level of rigor is consistent with standard practices in evaluating behavioral and attitudinal shifts in human-subject studies.

Our null hypothesis is as follows:
\begin{quote}
\textbf{H\textsubscript{0}: There is no significant difference in participants’ self-reported sustainability confidence, familiarity, or intentions before and after playing the VR game.}
\end{quote}

\textbf{Changes in intended behaviors} were drawn from two post-experience survey questions designed to capture participants’ self-reported intentions to adopt more sustainable habits after the VR experience.

\begin{itemize}
    \item Willingness to adopt specific food waste reduction practices at home (multiple-select).
    \item Perceived potential uses of food scraps, reflecting increased awareness of reuse possibilities (open-ended).
\end{itemize}

To evaluate changes in intended behaviors, we utilized two distinct methods:

\begin{itemize}
    \item For the \textbf{multi-select question}, we treated the answers as binary (whether or not they selected each practice), and compared pre- and post-survey responses using the Wilcoxon signed-rank test. This non-parametric test is appropriate as the data is paired (same individuals) and binary (whether a specific action was chosen or not). The significance threshold was set at 95\% (p < 0.05), as it is important to ensure the observed changes are not just due to chance but reflect a true shift in behavior intentions.
    \item For the \textbf{open-ended question}, we quantify the changes by grouping them into categories and calculate the percentage of participants who mentioned new reuse ideas in their post-survey responses that they did not mention in their pre-survey responses. This method provides insight into how the VR experience influenced participants’ awareness of food waste reduction strategies, particularly in terms of the creative reuse of food scraps.
\end{itemize}

Our null hypothesis for the Wilcoxon test is as follows: 

\begin{quote} 
\textbf{H\textsubscript{0}: There is no significant difference in the number of intended food waste reduction practices before and after playing the VR game.} 
\end{quote}

Similarly, for the open-ended question, the null hypothesis is that there will be no significant difference in the diversity of responses before and after the VR game. By comparing pre- and post-responses, we aim to identify whether the VR experience increased participants’ awareness of possible food waste reduction practices, as reflected in their intended actions.

For a complete list of survey questions and how they align with the metrics evaluated, refer to \textbf{Appendix ~\ref{appendix:pre_play_survey}} and \textbf{Appendix ~\ref{appendix:post_play_survey}}.

\section{Evaluation Results} 
\subsection{Results of Preliminary User Research}

The insights gathered from our preliminary user survey played a crucial role in shaping the direction and design of our climate change escape room. With 41 participants (90. 2\% being university students and 9.8\% being working adults) responding to questions about their familiarity with the solutions to climate change and the actions they take, we identified several key trends that influenced the structure of the escape room experience.

\subsubsection{General Familiarity with Climate Change Measures}

A 5-point scale was used to evaluate this, with 1 being "Not at all familiar" and 5 being "Very familiar". Of the 41 participants, 20 rated themselves moderately familiar (3 out of 5). 11 participants indicated that they were more familiar (ratings of 4 or 5), while the other 10 rated themselves at the lower end (1 or 2).

These results show that while most participants have a basic level of familiarity, there is room for further education on climate change solutions and their impact.

\subsubsection{Actions Taken to Combat Climate Change}

Participants were asked which climate-positive habits they already incorporate into their daily lives.

The most common actions reported are as follows.
\begin{itemize}
    \item Reducing energy usage (e.g., turning off appliances): 34 (82.9\%)
    \item Choosing public transport, cycling, or walking: 30 (73.2\%)
    \item Using reusable bags and bottles: 29 (70.7\%)
    \item Reducing food waste: 26 (63.4\%)
\end{itemize}

However, certain sustainable actions were less commonly practiced:

\begin{itemize}
    \item Choosing products with minimal packaging: 9 (22\%)
    \item Reducing meat consumption: 8 (19.5\%)
    \item Supporting sustainable businesses: 6 (14.6\%)
    \item Composting food waste: 2 (4.9\%)
\end{itemize}

\begin{figure}[H]
    \centering
    \includegraphics[width=\columnwidth]{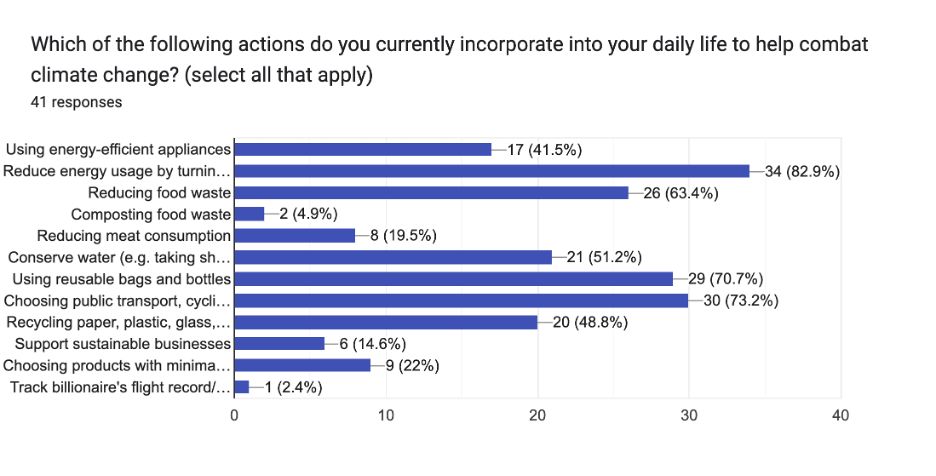}
    \caption{Current Actions Taken by Participants to Combat Climate Change}
\end{figure}

In response to these findings, the escape room was designed to focus on these least common habits. These are areas where engagement could be improved, and participants would benefit from learning about the impact of their choices while shopping and managing their food waste.

\subsubsection{Familiarity with Large-scale Climate Efforts}

Participants were asked how familiar they were with several large-scale climate initiatives. The breakdown of responses for each initiative was as follows:

\begin{figure}[H]
    \centering
    \includegraphics[width=\columnwidth]{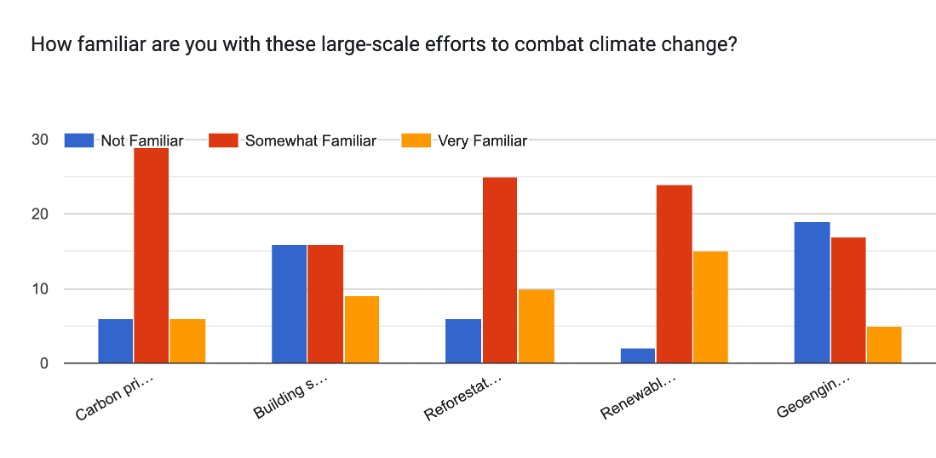}
    \caption{Familiarity with Large-scale Climate Efforts}
\end{figure}

The survey results showed that participants are most familiar with renewable energy and reforestation, but less familiar with complex strategies, such as geoengineering and carbon pricing policies.

\subsubsection{Awareness of Singapore-Specific Climate Policies} 

When asked about local climate-related policies in Singapore, the results showed varying levels of recognition.

\begin{figure}[H]
    \centering
    \includegraphics[width=\columnwidth]{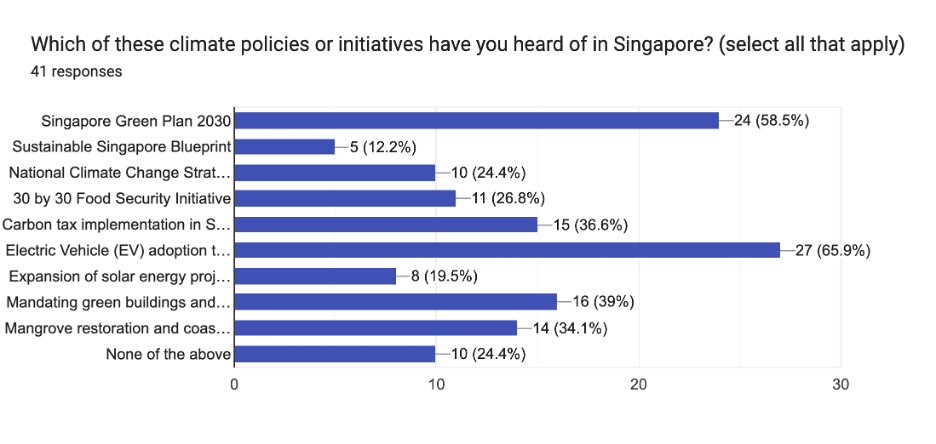}
    \caption{Participants' Awareness of Singapore-Specific Climate Policies}
\end{figure}

These findings highlight the recognition of major initiatives, such as the EV adoption targets and Green Plan 2030, while less awareness was found for more specific policies. In response to these results, the escape room design will emphasize Singapore-specific policies that are less recognized (such as the 30 by 30 Food Security Initiative), helping participants better understand Singapore's role in combating climate change.

\subsubsection{Conclusion and Design Decisions}
Based on the results of the preliminary user research, the escape room design was influenced by participants' familiarity with and engagement in sustainable practices. For the \textbf{Supermarket Room}, we focused on actions participants practiced the least, such as choosing products with minimal packaging, reducing meat consumption, and supporting sustainable businesses, aiming to raise awareness and encourage more conscious shopping.

In the \textbf{Food Waste Room/Kitchen}, we combined reducing food waste (63.4\%) with composting food waste (4.9\%) to create a cohesive learning experience that bridges familiar behaviors with lesser-known but impactful practices.

These results guide our escape room's design, addressing both popular and less-adopted sustainable practices to engage participants and foster commitment to climate action. In future iterations, we plan to incorporate more lesser-recognized policies, habits, and large-scale initiatives, which have not been included due to time and resource constraints. This will allow us to further broaden the educational scope and enhance the overall impact of the game.

\subsection{Results of Application Evaluation}

\subsubsection{Objective Knowledge Gains} 
Participants showed improved knowledge across all three evaluated topics:
\begin{itemize}
    \item \textbf{Environmental Impact:} Correct responses increased from 27.78\% to 41.67\%; Wilcoxon statistic = 28.0, \textit{p = 0.083}. \\
    \textit{This result is sufficient to reject the null hypothesis at the 90\% significance level.}
    
    \item \textbf{Food Packaging:} Correct responses increased from 44.44\% to 58.33\%; Wilcoxon statistic = 40.0, \textit{p = 0.098}. \\
    \textit{This result is sufficient to reject the null hypothesis at the 90\% significance level.}
    
    \item \textbf{Food Waste:} Correct responses increased from 61.11\% to 75.00\%; Wilcoxon statistic = 10.0, \textit{p = 0.048}. \\
    \textit{This result is sufficient to reject the null hypothesis at the 90\% significance level.}
\end{itemize}
A detailed breakdown of the evaluation method and Wilcoxon test statistics for each knowledge item is presented in Appendix~\ref{appendix:objective_knowledge_gains}.

These results suggest that the VR experience led to statistically significant improvements in participants’ ability to recognize sustainable practices in all three domains.

\subsubsection{Subjective Shifts in Perceptions}  
Statistically significant positive shifts were observed across all subjective measures:
\begin{itemize}
    \item \textbf{Familiarity with food sustainability:} Wilcoxon statistic = 22.5, \textit{p = 0.0001}. \\
    \textit{This result is sufficient to reject the null hypothesis at the 95\% significance level.}
    
    \item \textbf{Confidence in making sustainable food choices:} Wilcoxon statistic = 22.0, \textit{p < 0.00001}. \\
    \textit{This result is sufficient to reject the null hypothesis at the 95\% significance level.}
    
    \item \textbf{Likelihood of considering sustainability when shopping:} Wilcoxon statistic = 7.0, \textit{p < 0.00001}. \\
    \textit{This result is sufficient to reject the null hypothesis at the 95\% significance level.}
\end{itemize}

A detailed breakdown of the evaluation method and Wilcoxon test statistics for each knowledge item is presented in Appendix~\ref{appendix:subjective_perceptions}.

These findings indicate strong evidence that the VR experience significantly improved participants’ self-perceptions regarding sustainability.

\subsubsection{Changes in Intended Behaviors}
    A detailed breakdown of the analysis method used for this evaluation is provided in Appendix~\ref{appendix:intended_behaviors_analysis}.
\begin{itemize}
    \item \textbf{Results of Wilcoxon Test for Multi-Select Question}
    \begin{itemize}
        \item  The mean increase in the number of intended food waste reduction actions from pre- to post-survey is \textbf{1.64}. 
        \item  The Wilcoxon signed-rank test revealed a statistically significant increase: Wilcoxon statistic = 57.5, \textit{p = 0.00068}. \\
     \textit{This result is sufficient to reject the null hypothesis at the 95\% significance level.}
    \end{itemize}
    
    \item \textbf{Results of Analysis of Open-Ended Question}
    Pre-survey responses indicated that the vast majority of participants disposed of food scraps—such as vegetable peels, eggshells, and coffee grounds—by throwing them away, with only a small minority (12.12\%, 4 out of 33 valid pre-survey responses) mentioning any form of reuse, such as composting. Post-survey responses revealed a substantial shift ( 90\%, 27 out of 30 valid post-survey responses) mentioned at least one new reuse idea not previously stated, including composting, making protein or calcium powder from eggshells, using coffee grounds as cleaning agents, and repurposing scraps for DIY products like face masks or diffusers. This shift suggests a significant increase in awareness and creativity related to food scrap reuse.

\end{itemize}

These findings provide strong evidence that the VR experience significantly enhanced participants' intentions to adopt sustainable food-related behaviors.


\subsubsection{Limitations and Future Work}
\begin{enumerate}
    \item \textbf{Limitations} \\
    While our study has yielded positive results, we would like to acknowledge the following limitations:
    \begin{itemize}
        \item \textbf{Limited Sample Size}: The study involved 36 university students in Singapore, which limits the generalizability of our findings to a broader population.
        \item \textbf{Potential Short-term Impact}: We assessed immediate changes in knowledge, confidence, and behavioral intentions. The long-term effects of the VR experience on sustained behavior change remain unclear.
        \item \textbf{VR Accessibility}: The cost and accessibility of VR technology can be a barrier to wider adoption of this educational approach.
        \item \textbf{Narrow Scope}: The application focuses primarily on food-related sustainability; it does not cover the full spectrum of sustainable living practices.
        \item \textbf{Assessment Methods}: Our study relied primarily on self-reported measures. More objective measures of behavior change could provide additional insights.
        \item \textbf{Significance Threshold}: Our use of a 90\% significance level (p < 0.10) for analyzing knowledge gains increases the risk of Type I errors (false positives).
    \end{itemize}

    \item \textbf{Future Work} \\
    \begin{itemize}
        \item \textbf{Larger and More Diverse Samples}: Replicating the study with larger, more diverse participant groups, including individuals from various age groups and backgrounds, will improve the generalizability of the findings.
        \item \textbf{Longitudinal Studies}: Conducting longitudinal studies to track behavior change over extended periods will provide a better understanding of the long-term effectiveness of the VR experience and how well intentions translate into action.
        \item \textbf{Objective Behavior Measures}: Incorporating objective measures of behavior change, such as tracking purchasing decisions, waste generation, or energy consumption, will provide a more accurate assessment of the VR experience's impact on actual behavior.
        \item \textbf{Accessible Platforms}: Exploring the use of more accessible technologies, such as mobile VR or augmented reality (AR), can broaden the reach of this educational tool.
        \item \textbf{Expanded Scope}: Developing additional VR scenarios that cover a wider range of sustainability topics, including energy conservation, transportation, and consumption habits, will provide a more comprehensive educational experience.
        \item \textbf{Standard Significance Levels}: Future studies should strive to use the standard 95\% significance level (p < 0.05) to minimize the risk of false positives, especially with larger sample sizes.
    \end{itemize}
\end{enumerate}

\section{Acknowledgments}
We would like to thank our professor, Professor Anand Bhojan, Senior Lecturer at the National University of Singapore, for their invaluable guidance and feedback throughout this research project. We also extend our gratitude to Chou Chih-An, Master's student in Computer Science at the National University of Singapore, for her contribution during the ideation phase, which helped shape the initial direction of our work. Additionally, we would like to acknowledge the support and assistance provided by our teaching assistants, Xu Peisen and Zhao Yiran, both PhD students in Computer Science at the National University of Singapore, whose feedback and guidance were instrumental in refining our project.


\section{Conclusion}

In this paper, we presented Meltdown, an innovative VR escape room application designed to tackle the perception gap in climate change education.  By simulating real-world scenarios of grocery shopping and food waste management, our application effectively immerses users in the consequences of their consumption choices.  The results of our user study are highly encouraging, demonstrating significant improvements in participants' knowledge, confidence, and intention to adopt sustainable behaviors.  We believe that Meltdown offers a powerful and engaging tool for promoting climate literacy and fostering a deeper connection between individual actions and environmental impact.  This project showcases the immense potential of VR technology to transform educational experiences and inspire positive change towards a more sustainable future. 



\bibliographystyle{ACM-Reference-Format}
\bibliography{sample-base}


\begin{thebibliography}{15}


\ifx \showCODEN    \undefined \def \showCODEN     #1{\unskip}     \fi
\ifx \showDOI      \undefined \def \showDOI       #1{#1}\fi
\ifx \showISBNx    \undefined \def \showISBNx     #1{\unskip}     \fi
\ifx \showISBNxiii \undefined \def \showISBNxiii  #1{\unskip}     \fi
\ifx \showISSN     \undefined \def \showISSN      #1{\unskip}     \fi
\ifx \showLCCN     \undefined \def \showLCCN      #1{\unskip}     \fi
\ifx \shownote     \undefined \def \shownote      #1{#1}          \fi
\ifx \showarticletitle \undefined \def \showarticletitle #1{#1}   \fi
\ifx \showURL      \undefined \def \showURL       {\relax}        \fi
\providecommand\bibfield[2]{#2}
\providecommand\bibinfo[2]{#2}
\providecommand\natexlab[1]{#1}
\providecommand\showeprint[2][]{arXiv:#2}

\bibitem[Grewe et~al\mbox{.}(2013)]%
        {grewe_2013}
\bibfield{author}{\bibinfo{person}{Philip Grewe}, \bibinfo{person}{Agnes Kohsik}, \bibinfo{person}{David Flentge}, \bibinfo{person}{Eugen Dyck}, \bibinfo{person}{Mario Botsch}, \bibinfo{person}{York Winter}, \bibinfo{person}{Hans~J Markowitsch}, \bibinfo{person}{Christian~G Bien}, {and} \bibinfo{person}{Martina Piefke}.} \bibinfo{year}{2013}\natexlab{}.
\newblock \showarticletitle{Learning real-life cognitive abilities in a novel 360°-virtual reality supermarket: a neuropsychological study of healthy participants and patients with epilepsy}.
\newblock \bibinfo{journal}{\emph{Journal of Neuroengineering and Rehabilitation}} \bibinfo{volume}{10}, \bibinfo{number}{1} (\bibinfo{date}{Apr} \bibinfo{year}{2013}), \bibinfo{pages}{42–42}.
\newblock
\urldef\tempurl%
\url{https://doi.org/10.1186/1743-0003-10-42}
\showDOI{\tempurl}


\bibitem[Group(2023)]%
        {fairprice_group_2023}
\bibfield{author}{\bibinfo{person}{FairPrice Group}.} \bibinfo{year}{2023}\natexlab{}.
\newblock \bibinfo{title}{NTUC FairPrice saves over 57 million plastic bags in 2022, surpassing and nearly doubling annual target - FairPrice Group}.
\newblock
\newblock
\urldef\tempurl%
\url{https://www.fairpricegroup.com.sg/ntuc-fairprice-saves-over-57-million-plastic-bags-in-2022-surpassing-and-nearly-doubling-annual-target/}
\showURL{%
\tempurl}


\bibitem[Guide(2021)]%
        {scholarship_guide_2021}
\bibfield{author}{\bibinfo{person}{Scholarship Guide}.} \bibinfo{year}{2021}\natexlab{}.
\newblock \bibinfo{title}{This App Will Help You Live a More Sustainable Life: susGain}.
\newblock
\newblock
\urldef\tempurl%
\url{https://blog.scholarshipguide.com.sg/smart-sustainable-habits-worth-cultivating-introducing-susgain/}
\showURL{%
\tempurl}


\bibitem[Huang et~al\mbox{.}(2021)]%
        {Huang03062021}
\bibfield{author}{\bibinfo{person}{Jiawei Huang}, \bibinfo{person}{Melissa~S. Lucash}, \bibinfo{person}{Robert~M. Scheller}, {and} \bibinfo{person}{Alexander~Klippel and}.} \bibinfo{year}{2021}\natexlab{}.
\newblock \showarticletitle{Walking through the forests of the future: using data-driven virtual reality to visualize forests under climate change}.
\newblock \bibinfo{journal}{\emph{International Journal of Geographical Information Science}} \bibinfo{volume}{35}, \bibinfo{number}{6} (\bibinfo{year}{2021}), \bibinfo{pages}{1155--1178}.
\newblock
\urldef\tempurl%
\url{https://doi.org/10.1080/13658816.2020.1830997}
\showDOI{\tempurl}
\showeprint{https://doi.org/10.1080/13658816.2020.1830997}


\bibitem[Kolaczkowski(2021)]%
        {MaciejKolaczkowski2021}
\bibfield{author}{\bibinfo{person}{Maciej Kolaczkowski}.} \bibinfo{year}{2021}\natexlab{}.
\newblock \bibinfo{title}{What can consumers do to help solve the climate change crisis?}
\newblock
\newblock
\urldef\tempurl%
\url{https://www.weforum.org/stories/2021/02/consumers-help-solve-climate-change/}
\showURL{%
\tempurl}


\bibitem[Lo and Tsai(2022)]%
        {lo_tsai_2022}
\bibfield{author}{\bibinfo{person}{Shih-Che Lo} {and} \bibinfo{person}{Hung-Hsu Tsai}.} \bibinfo{year}{2022}\natexlab{}.
\newblock \showarticletitle{Design of 3D Virtual Reality in the Metaverse for Environmental Conservation Education Based on Cognitive Theory}.
\newblock \bibinfo{journal}{\emph{Sensors (14248220)}} \bibinfo{volume}{22}, \bibinfo{number}{21} (\bibinfo{date}{Nov} \bibinfo{year}{2022}), \bibinfo{pages}{8329}.
\newblock
\urldef\tempurl%
\url{https://doi.org/10.3390/s22218329}
\showDOI{\tempurl}


\bibitem[NEA(2023)]%
        {nea_2023}
\bibfield{author}{\bibinfo{person}{NEA}.} \bibinfo{year}{2023}\natexlab{}.
\newblock \bibinfo{title}{Waste statistics and overall recycling}.
\newblock
\newblock
\urldef\tempurl%
\url{https://www.nea.gov.sg/our-services/waste-management/waste-statistics-and-overall-recycling}
\showURL{%
\tempurl}


\bibitem[Novo et~al\mbox{.}(2024)]%
        {novo_zanchetta_goldmann_vaz_2024}
\bibfield{author}{\bibinfo{person}{Carolina Novo}, \bibinfo{person}{Chiara Zanchetta}, \bibinfo{person}{Elisa Goldmann}, {and} \bibinfo{person}{Carlos Vaz}.} \bibinfo{year}{2024}\natexlab{}.
\newblock \showarticletitle{The Use of Gamification and Web-Based Apps for Sustainability Education}.
\newblock \bibinfo{journal}{\emph{Sustainability}} \bibinfo{volume}{16}, \bibinfo{number}{8} (\bibinfo{date}{Apr} \bibinfo{year}{2024}), \bibinfo{pages}{3197–3197}.
\newblock
\urldef\tempurl%
\url{https://doi.org/10.3390/su16083197}
\showDOI{\tempurl}


\bibitem[of~Education(2025)]%
        {moe_2025}
\bibfield{author}{\bibinfo{person}{Ministry of Education}.} \bibinfo{year}{2025}\natexlab{}.
\newblock \bibinfo{title}{Sustainable Living @ South East}.
\newblock
\newblock
\urldef\tempurl%
\url{https://necrg.moe.edu.sg/workshops-for-students/south-east-cdc/sustainable-living-se/}
\showURL{%
\tempurl}


\bibitem[of~Sustainability and the Environment(2024)]%
        {mse_2024}
\bibfield{author}{\bibinfo{person}{Ministry of Sustainability} {and} \bibinfo{person}{the Environment}.} \bibinfo{year}{2024}\natexlab{}.
\newblock \bibinfo{title}{Oral Reply to Parliamentary Question on Singapore Green Labelling Scheme}.
\newblock
\newblock
\urldef\tempurl%
\url{https://www.mse.gov.sg/latest-news/oral-reply-to-pq-on-singapore-green-labelling-scheme}
\showURL{%
\tempurl}


\bibitem[Onita et~al\mbox{.}(2025)]%
        {cristiana_2025}
\bibfield{author}{\bibinfo{person}{Cristiana~Amalia Onita}, \bibinfo{person}{Daniela-Viorelia Matei}, \bibinfo{person}{Ilie Onu}, \bibinfo{person}{Daniel-Andrei Iordan}, \bibinfo{person}{Elena Chelarasu}, \bibinfo{person}{Nicoleta Tupita}, \bibinfo{person}{Diana Petrescu-Miron}, \bibinfo{person}{Mihaela Radeanu}, \bibinfo{person}{Georgiana Juravle}, \bibinfo{person}{Calin Corciova}, \bibinfo{person}{Robert Fuior}, {and} \bibinfo{person}{Veronica Mocanu}.} \bibinfo{year}{2025}\natexlab{}.
\newblock \showarticletitle{Using VR Supermarket for Nutritional Research and Education: A Scoping Review}.
\newblock \bibinfo{journal}{\emph{Nutrients}} \bibinfo{volume}{17}, \bibinfo{number}{6} (\bibinfo{date}{Mar} \bibinfo{year}{2025}), \bibinfo{pages}{999–999}.
\newblock
\urldef\tempurl%
\url{https://doi.org/10.3390/nu17060999}
\showDOI{\tempurl}


\bibitem[Tan and Nurul‐Asna(2023)]%
        {tan_nurul‐asna_2023}
\bibfield{author}{\bibinfo{person}{Cedric K.~W. Tan} {and} \bibinfo{person}{Hidayah Nurul‐Asna}.} \bibinfo{year}{2023}\natexlab{}.
\newblock \showarticletitle{Serious games for environmental education}.
\newblock \bibinfo{journal}{\emph{Integrative Conservation}} \bibinfo{volume}{2}, \bibinfo{number}{1} (\bibinfo{date}{Mar} \bibinfo{year}{2023}), \bibinfo{pages}{19–42}.
\newblock
\urldef\tempurl%
\url{https://doi.org/10.1002/inc3.18}
\showDOI{\tempurl}


\bibitem[Tan(2020)]%
        {tan_2020}
\bibfield{author}{\bibinfo{person}{Judith Tan}.} \bibinfo{year}{2020}\natexlab{}.
\newblock \bibinfo{title}{Scaling towards zero food waste}.
\newblock
\newblock
\urldef\tempurl%
\url{https://www.straitstimes.com/singapore/scaling-towards-zero-food-waste}
\showURL{%
\tempurl}


\bibitem[Xu et~al\mbox{.}(2022)]%
        {9815983}
\bibfield{author}{\bibinfo{person}{Zixiang Xu}, \bibinfo{person}{Yuan Liang}, \bibinfo{person}{Abraham~G. Campbell}, {and} \bibinfo{person}{Soumyabrata Dev}.} \bibinfo{year}{2022}\natexlab{}.
\newblock \showarticletitle{An Explore of Virtual Reality for Awareness of the Climate Change Crisis: A Simulation of Sea Level Rise}. In \bibinfo{booktitle}{\emph{2022 8th International Conference of the Immersive Learning Research Network (iLRN)}}. \bibinfo{pages}{1--5}.
\newblock
\urldef\tempurl%
\url{https://doi.org/10.23919/iLRN55037.2022.9815983}
\showDOI{\tempurl}


\bibitem[Yuwono et~al\mbox{.}({[n.\,d.]})]%
        {Yuwono_Chong_Harijanto_Xu}
\bibfield{author}{\bibinfo{person}{Florentiana Yuwono}, \bibinfo{person}{Xiao~Xuan Chong}, \bibinfo{person}{Melissa~Anastasia Harijanto}, {and} \bibinfo{person}{Yi Xu}.} \bibinfo{year}{[n.\,d.]}\natexlab{}.
\newblock \bibinfo{title}{Meltdown GitHub Repository}.
\newblock
\newblock
\urldef\tempurl%
\url{https://github.com/florentianayuwono/meltdown}
\showURL{%
\tempurl}


\end{thebibliography}

\clearpage
\appendix
\section{Preliminary User Research Questions}
\label{appendix:preliminary_user_research_qns}
\begin{enumerate}
    \item Occupation
    \begin{itemize}
        \item Primary to JC level student
        \item University student
        \item Working adult
        \item Retiree
        \item Unemployed
    \end{itemize}
    \item How familiar would you consider yourself with measures to combat climate change?
    \begin{itemize}
        \item Rate on a scale of 1 to 5, with 1 being "Not at all familiar", and 5 being "Very familiar".
    \end{itemize}
    \item Which of the following actions do you currently incorporate into your daily life to help combat climate change? (select all that apply)
    \begin{itemize}
        \item Using energy-efficient appliances
        \item Reduce energy usage by turning off appliances when not in use
        \item Reducing food waste
        \item Composting food waste
        \item Reducing meat consumption
        \item Conserve water (e.g. taking shorter showers, don't leave tap running)
        \item Using reusable bags and bottles
        \item Choosing public transport, cycling, or walking over private cars
        \item Recycling paper, plastic, glass, and other recyclable materials
        \item Support sustainable businesses
        \item Choosing products with minimal packaging
    \end{itemize}
    \item How familiar are you with these large-scale efforts to combat climate change?
    \begin{itemize}
        \item Carbon pricing policies (e.g., carbon tax, cap-and-trade systems) 
        \begin{itemize}
            \item Not Familiar
            \item Somewhat Familiar
            \item Very Familiar
        \end{itemize}
        \item Building sea walls to protect against rising sea levels
        \begin{itemize}
            \item Not Familiar
            \item Somewhat Familiar
            \item Very Familiar
        \end{itemize}
        \item Reforestation and afforestation projects
        \begin{itemize}
            \item Not Familiar
            \item Somewhat Familiar
            \item Very Familiar
        \end{itemize}
        \item Renewable energy investments (solar, wind, hydro)
        \begin{itemize}
            \item Not Familiar
            \item Somewhat Familiar
            \item Very Familiar
        \end{itemize}
        \item Geoengineering solutions (e.g., cloud seeding, carbon capture)
        \begin{itemize}
            \item Not Familiar
            \item Somewhat Familiar
            \item Very Familiar
        \end{itemize}
    \end{itemize}
    \item Which of these climate policies or initiatives have you heard of in Singapore? (select all that apply)
    \begin{itemize}
        \item Singapore Green Plan 2030
        \item Sustainable Singapore Blueprint
        \item National Climate Change Strategy
        \item 30 by 30 Food Security Initiative
        \item Carbon tax implementation in Singapore
        \item Electric Vehicle (EV) adoption targets and incentives
        \item Expansion of solar energy projects
        \item Mandating green buildings and sustainable urban planning
        \item Mangrove restoration and coastal protection efforts
        \item None of the above
    \end{itemize}
\end{enumerate}

\section{Pre-Play Survey}
\label{appendix:pre_play_survey}
\begin{enumerate}
    \item How familiar are you with the concept of food sustainability?
    \begin{itemize}
        \item Rate on a scale of 1 to 5, with 1 being "Not at all familiar", and 5 being "Very familiar".
    \end{itemize}
    \item Which factors contribute most to the environmental impact of food production? (Select all that apply)
    \begin{itemize}
        \item Farming methods
        \item The popularity of the food
        \item Food waste
        \item Transportation
        \item How expensive the food is
        \item Packaging
        \item Energy consumption
    \end{itemize}
    \item What are some characteristics of sustainable food packaging? (Select all that apply)
    \begin{itemize}
        \item Biodegradable materials
        \item Recyclable packaging
        \item Single-use plastics
        \item Uses less plastic
        \item Made from renewable resources
        \item More expensive packaging
        \item Not sure
    \end{itemize}
    \item Why is reducing food waste important for the environment? 
    \begin{itemize}
        \item Reduces greenhouse gas emissions
        \item Conserve resources like water and energy
        \item Lowers the demand for excessive food production
        \item Allows people to buy more food
        \item Makes food taste better
        \item Not sure
    \end{itemize}
    \item What actions do you currently take to reduce food waste at home? (Select all that apply)
    \begin{itemize}
        \item Meal planning to reduce excess food
        \item Using leftovers creatively
        \item Composting food scraps
        \item Donating surplus food
        \item Consume food before expiry dates
        \item I don’t do anything specific yet
    \end{itemize}
    \item  How do you usually dispose of food scraps like vegetable peels, eggshells, or coffee grounds?  
    \begin{itemize}
        \item Short answer, open-ended question.
    \end{itemize}
    \item How confident are you in making environmentally-friendly food choices when grocery shopping?  
    \begin{itemize}
        \item Rate on a scale of 1 to 5, with 1 being "Not confident at all", and 5 being "Very confident".
    \end{itemize}
    \item How likely are you to consider sustainability when shopping for food?  
    \begin{itemize}
        \item Rate on a scale of 1 to 5, with 1 being "Not likely at all", and 5 being "Very likely".
    \end{itemize}
\end{enumerate}
\section{Post-Play Survey}
\label{appendix:post_play_survey}
\begin{enumerate}
    \item To what extent do you feel that your concept of food sustainability has improved after this experience?
    \begin{itemize}
        \item Rate on a scale of 1 to 5, with 1 being "Did not improve at all", and 5 being "Improve greatly".
    \end{itemize}
    \item Which factors contribute most to the environmental impact of food production? (Select all that apply)
    \begin{itemize}
        \item Farming methods
        \item The popularity of the food
        \item Food waste
        \item Transportation
        \item How expensive the food is
        \item Packaging
        \item Energy consumption
    \end{itemize}
    \item What are some characteristics of sustainable food packaging? (Select all that apply)
    \begin{itemize}
        \item Biodegradable materials
        \item Recyclable packaging
        \item Single-use plastics
        \item Uses less plastic
        \item Made from renewable resources
        \item More expensive packaging
        \item Not sure
    \end{itemize}
    \item Why is reducing food waste important for the environment? 
    \begin{itemize}
        \item Reduces greenhouse gas emissions
        \item Conserve resources like water and energy
        \item Lowers the demand for excessive food production
        \item Allows people to buy more food
        \item Makes food taste better
        \item Not sure
    \end{itemize}
    \item What new actions would you consider taking to reduce food waste at home after this experience? (Select all that apply)
    \begin{itemize}
        \item Meal planning to reduce excess food
        \item Using leftovers creatively
        \item Composting food scraps
        \item Donating surplus food
        \item Consume food before expiry dates
        \item I don’t plan to change my habits
    \end{itemize}
    \item How do you now think food scraps like vegetable peels, eggshells, or coffee grounds can be used?
    \begin{itemize}
        \item Short answer, open-ended question.
    \end{itemize}
    \item To what extent has your confidence improved in making environmentally friendly food choices when shopping after this experience?
    \begin{itemize}
        \item Rate on a scale of 1 to 5, with 1 being "No improvement at all", and 5 being "Greatly improved".
    \end{itemize}
    \item How likely are you to consider sustainability when shopping for food after this experience?
    \begin{itemize}
        \item Rate on a scale of 1 to 5, with 1 being "Not likely at all", and 5 being "Very likely".
    \end{itemize}
\end{enumerate}
\section{User Testing Guide}
\label{appendix:user_testing_guide}
\subsubsection*{\textbf{Opening the session}}

\texttt{[Invite the participant to sit down]}

Thank you for agreeing to participate in our usability study. My name is \texttt{[facilitator’s name]}, and I will be facilitating the session today.

We are testing a VR escape room experience, and your feedback will help us improve it. The session will take approximately 30–40 minutes.

Important Notes:
The game will provide audio instructions upon entering each room. I will remain silent during this time to avoid interruptions.
I will speak to you only before you enter a room and after you complete its tasks, before proceeding to the next one.
If you feel unwell at any point or wish to stop, please let us know.
There are no right or wrong answers—we are testing the experience, not you.

Do you have any questions before we begin?
Before we begin, may I seek your consent to record this session for research purposes?

If hesitant: Recording is used for our personal reference purposes in case we miss out on anything that was said during the session itself. It will not be shared anywhere else and everything shared in this session will be kept confidential on our part and be used solely for the purposes of research. 

Notetaker STARTS: 
\begin{itemize}
    \item Phone screen recording
    \item Audio recording
\end{itemize}

\subsubsection*{\textbf{Pre-test Questions (5 min)}}

Before we start, I will be asking some questions to get to know you better. 
\begin{enumerate}
    \item What is your occupation? (Working/ Student)
    \item Have you used VR before?
    \item (If user has used VR) What types of VR equipment/applications have you used?
    \item (If user has used VR) Do you experience motion sickness when using VR?
    \item What are your expectations of a VR escape room experience?
\end{enumerate}

\subsubsection*{\textbf{Scenarios (15 min)}}
General Instructions:
\begin{itemize}
    \item You can now stand up, wear the VR headset and begin the experience.
    \item You will enter the first room and listen to the in-game announcement providing instructions.
    \item Once the announcement ends, you may begin interacting with the environment.
    \item I will observe but will not interact unless you have finished the room or requested for help.
    \item Remember to think out loud as you play. For example, you can say what you see, what you think it will do, what you are going to do with it, etc.
\end{itemize}

\subsubsection*{\textbf{Task 1: Navigate the tutorial room}}
You have just entered the game and find yourself in this room. Navigate around it and show me what you would do. Let me know when you think you are done.

Success looks like: User is able to interact with everything in the tutorial room.

Questions:
\begin{itemize}
    \item What do you think this room is about?
    \item How did you know what to do?
    \item Was there anything unclear or confusing?
    \item How easy or difficult was it to interact with objects?
    \item Did you experience any discomfort or confusion when moving around?
\end{itemize}

\subsubsection*{\textbf{Task 2a: Enter the supermarket and complete the puzzle.}}
Now that you are done with the first room, you can go ahead and enter the second room. Follow the instructions and interact with the room to complete the puzzle. Let me know when you think you are done.
\begin{itemize}
    \item How did you feel about the transition between rooms?
    \item What do you think this room is about?
    \item Was there anything unclear or confusing?
    \item What stood out to you most in the room?
\end{itemize}

\subsubsection*{\textbf{Task 2b: Enter the kitchen and complete the puzzle}}
Now that you are done with the second room, you can go ahead and enter the third room. Follow the instructions and interact with the room to complete the puzzle. Let me know when you think you are done.
\begin{itemize}
    \item What do you think this room is about?
    \item Was there anything unclear or confusing?
    \item What stood out to you most in the room?
\end{itemize}

\subsubsection*{\textbf{Task 3: Enter the last room to complete the game}}
You have escaped the rooms, you may now enter the final room and complete the game.
\begin{itemize}
    \item What do you think of the ending of the game?
    \item Was there anything unclear or confusing?
\end{itemize}

\subsubsection*{\textbf{Post-test Questions (10 min)}}

Thank you. We also have some other questions to ask you. Please answer them honestly:
\begin{itemize}
    \item On a scale of 1–5, 1 being very difficult and 5 being very easy, how easy was it to understand and complete the escape room tasks?
    \item On a scale of 1–5, 1 being not comfortable at all and 5 being very comfortable, how comfortable was the movement and navigation?
    \item Did you feel immersed in the experience? What did not make/made you feel immersed?
    \item What was your favorite part of the experience?
    \item What was the most difficult or frustrating part?\item Would you play this escape room again? Why or why not?
    \item If you could change one thing about the experience, what would it be?
    \item Do you have any other feedback or suggestions?
\end{itemize}

\subsubsection*{\textbf{Closing}}
Do you \texttt{[participant]} have any last comments or questions for us? That’s all for the session today, thank you so much for your time!

\section{Detailed Statistical Results}
\label{appendix:results}

\subsection{Objective Knowledge Gains}
\label{appendix:objective_knowledge_gains}

Each objective knowledge question in the survey was a multiple-select item with a set of correct options. To ensure rigor in assessment, a response was counted as \textbf{correct only if the participant selected \emph{all} of the correct options} and no others. Partial selections were not awarded credit.

This binary scoring method (correct = 1, incorrect = 0) was applied to paired pre- and post-experience data. For each of the three questions, we computed the total number of correct responses and the corresponding percentage. These scores were then evaluated using the \textbf{Wilcoxon signed-rank test (one-sided)} to determine whether there was a statistically significant increase in the proportion of fully correct responses after the VR experience.

The full implementation logic was automated via a Python script, which calculates the Wilcoxon statistic and p-value of the test. The script also generates the distribution graph of pre- and post- responses, calculates the percentage of correct answers for both responses.:
\begin{enumerate}
    \item Load the pre- and post-experience survey CSV files into data frames.
    \item Ensure that column names are stripped of leading/trailing whitespace.
    \item Confirm that the number of rows (i.e., participants) match between the two datasets.
    \item Define a mapping of question identifiers to their corresponding column headers in the dataset.
    \item Specify the set of correct answers for each question.
    \item For each question:
    \begin{enumerate}
        \item Define a scoring function that returns 1 if \emph{all} correct answers (and only those) were selected by a participant; otherwise return 0.
        \item Apply this function to both the pre- and post-survey responses to generate binary correctness indicators.
        \item Compute the total number and percentage of correct responses before and after the VR experience, which are given by:
            \begin{equation*}
            \frac{\text{Sum of Correct Responses}}{\text{Total Responses}} \times 100
            \end{equation*}
        \item Conduct a one-sided Wilcoxon signed-rank test to evaluate whether post-survey scores are significantly higher than pre-survey scores.
    \end{enumerate}
    \item For each question, count how many times each answer option was selected in both pre- and post-surveys.
    \item Generate one bar plot per question comparing pre- and post-experience option selection frequencies.
    \begin{enumerate}
        \item Correct options are visually highlighted in green to indicate the ground truth.
        \item All graphs are displayed separately for clearer comparison.
    \end{enumerate}
\end{enumerate}

Due to the exploratory nature of this study and the limited sample size, we adopted a more lenient significance threshold of \textbf{p < 0.10} to detect meaningful trends that might otherwise be obscured.

\begin{table}[H]
\centering
\resizebox{\columnwidth}{!}{%
\begin{tabular}{|l|c|c|c|c|}
\hline
\textbf{Topic} & \textbf{Pre Correct (\%)} & \textbf{Post Correct (\%)} & \textbf{Wilcoxon Statistic} & \textbf{P-value} \\
\hline
Environmental Impact & 27.78\% & 41.67\% & 28.0 & 0.083 \\
Food Packaging & 44.44\% & 58.33\% & 40.0 & 0.098 \\
Food Waste & 61.11\% & 75.00\% & 10.0 & 0.048 \\
\hline
\end{tabular}%
}
\caption{Wilcoxon test results for objective knowledge-based questions.}
\end{table}

\begin{figure}[H]
        \centering
        \includegraphics[width=\columnwidth]{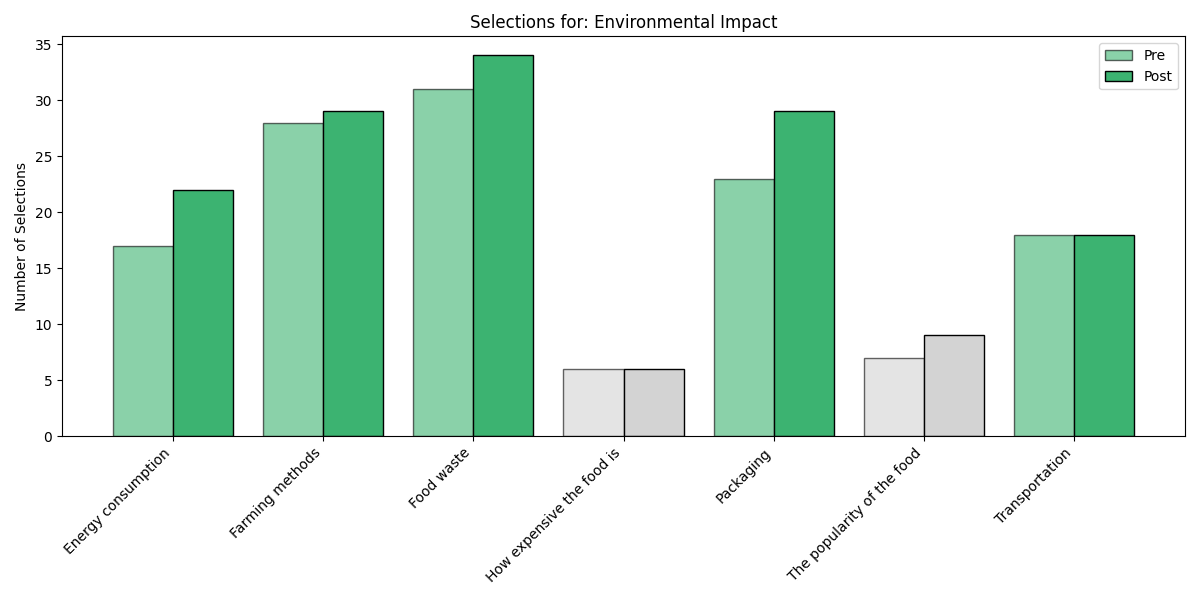}
        \caption{Participant Selections: Environmental Impact Factors Before and After VR Experience}
\end{figure}
\begin{figure}[H]
        \centering
        \includegraphics[width=\columnwidth]{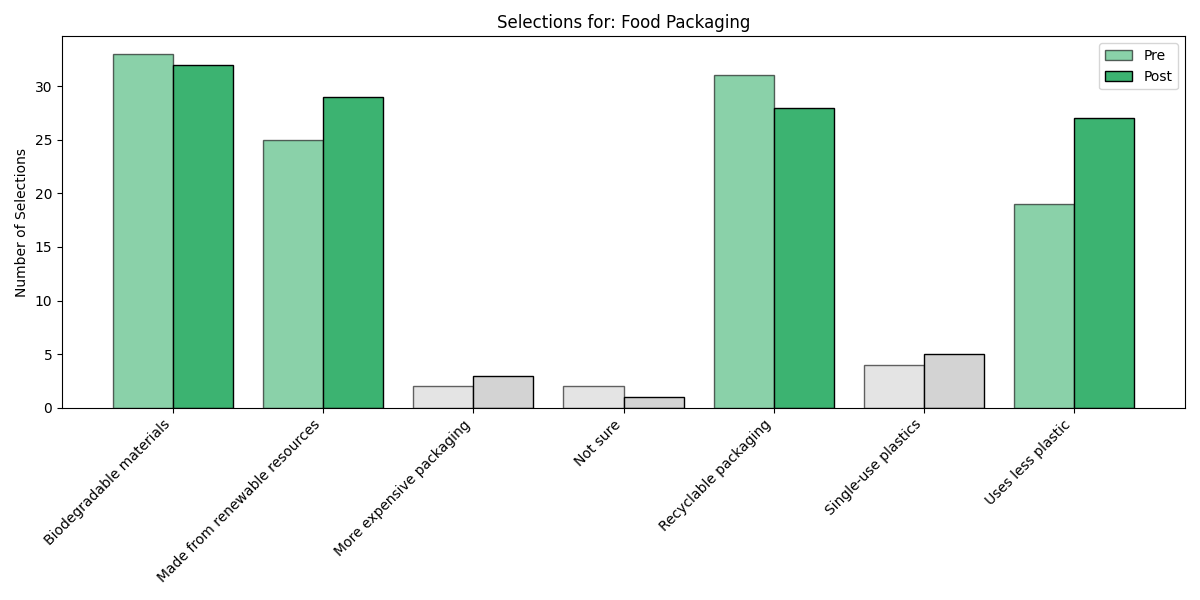}
        \caption{Participant Selections: Characteristics of Sustainable Food Packaging Before and After VR Experience}
\end{figure}
\begin{figure}[H]
        \centering
        \includegraphics[width=\columnwidth]{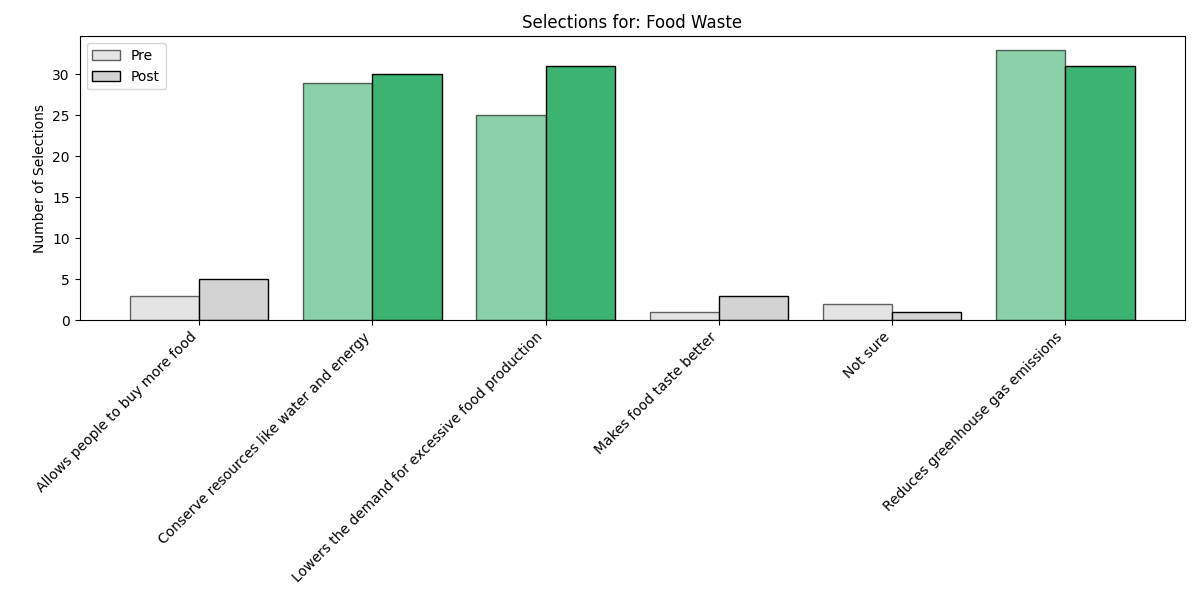}
        \caption{Participant Selections: Importance of Reducing Food Waste Before and After VR Experience}
\end{figure}

\subsection{Subjective Perception and Confidence Shifts}
\label{appendix:subjective_perceptions}

To analyze subjective shifts in participants’ self-reported familiarity, confidence, and intentions related to sustainability, we compared responses from the pre- and post-surveys. Each item was rated on a Likert scale (i.e., 1–5), and the differences in participants’ responses before and after the VR experience were evaluated using the \textbf{Wilcoxon signed-rank test (two-sided)}.

This analysis was conducted using a Python script that calculates the Wilcoxon statistic and p-value. It also generates graphs of the distributions of responses. The pseudocode is listed below:
\begin{enumerate}
    \item Loaded the pre- and post-survey datasets containing Likert-scale responses.
    \item Cleaned the column names by stripping whitespace.
    \item Converted all survey entries to numeric values, coercing non-numeric entries to NaN.
    \item Filled missing values by imputing the mean of each column.
    \item Paired each pre-survey question with its corresponding post-survey version.
    \item Applied the Wilcoxon signed-rank test to each paired dataset to evaluate whether there was a statistically significant difference in self-reported scores before and after the intervention.
    \item Prepared the data for visualization by tagging responses with their corresponding survey timepoint (Pre or Post).
    \item Plotted separate boxplots for each subjective question to visualize the distribution of responses before and after the experience.
\end{enumerate}

A significance level of \textbf{p < 0.05} was used to assess whether the observed changes in participants’ perceptions and intentions were unlikely to be due to chance.

\begin{table}[H]
\centering
\resizebox{\columnwidth}{!}{%
\begin{tabular}{|l|c|c|}
\hline
\textbf{Measure} & \textbf{Wilcoxon Statistic} & \textbf{P-value} \\
\hline
Familiarity with food sustainability & 22.5 & 0.0001 \\
Confidence in sustainable food choices & 22.0 & < 0.00001 \\
Likelihood of considering sustainability when shopping & 7.0 & < 0.00001 \\
\hline
\end{tabular}%
}
\caption{Wilcoxon test results for subjective sustainability-related questions.}
\end{table}

\begin{figure}[H]
        \centering
        \includegraphics[width=\columnwidth]{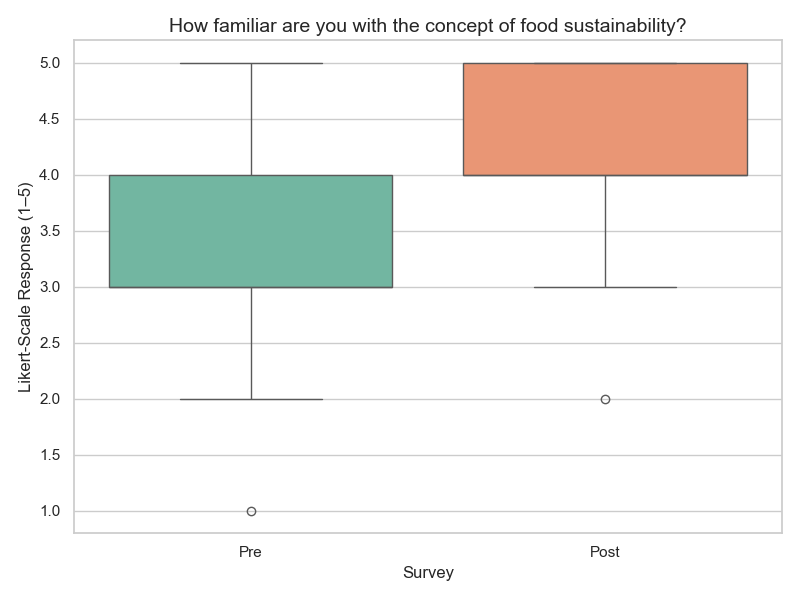}
        \caption{Distribution of Responses: How familiar are you with the concept of food sustainability?}
\end{figure}
\begin{figure}[H]
        \centering
        \includegraphics[width=\columnwidth]{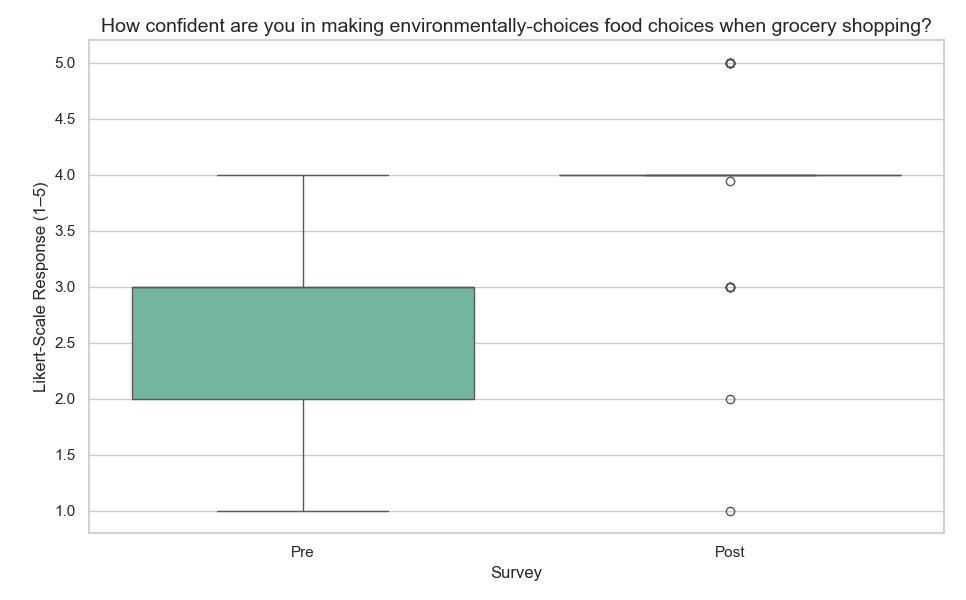}
        \caption{Distribution of Responses: How confident are you in making environmentally-choices food choices when grocery shopping?}
\end{figure}
\begin{figure}[H]
        \centering
        \includegraphics[width=\columnwidth]{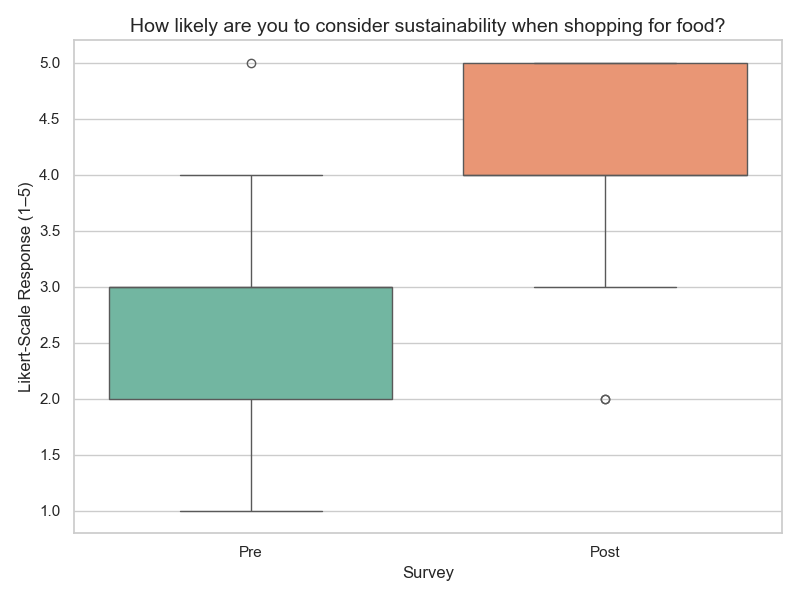}
        \caption{Distribution of Responses: How likely are you to consider sustainability when shopping for food?}
\end{figure}

\subsection{Changes in Intended Behaviors} \label{appendix:intended_behaviors_analysis}

To evaluate changes in participants’ intended behaviors regarding sustainability, we analyzed two survey items: one assessing willingness to adopt specific food waste reduction actions and another assessing awareness of the potential reuse of food scraps. The survey responses were provided in a multiple-select format for actions and an open-ended format for reuse possibilities.
\begin{enumerate}
    \item For the multi-select question, the full implementation logic was automated via a Python script that implements the Wilcoxon test on our survey results. The distribution of intended actions for both pre- and post-surveys are also generated via this script.

    The pseudocode for the process is as follows:
    \begin{enumerate}
        \item Load pre- and post-survey CSV data.
        \item Clean column names by stripping extra spaces.
        \item Parse answers from the pre-survey (convert to pre-action set, if participants answer "I don’t do anything specific yet", then they are given a an empty pre-action set).
        \item Parse answers from the post-survey (convert to set, if participants answer "I don’t plan to change my habits", they are given the same set of actions as their pre-action set).
        \item Calculate the set difference (new actions) between post and pre.
        \item Store the number of new actions, pre-actions, and post-actions for each participant.
        \item Perform Wilcoxon signed-rank test on pre- and post-action counts
        \item Output descriptive statistics (mean, median, etc.) and Wilcoxon test results
        \item Visualize the comparison with a boxplot.
        \item Interpret the statistical results: If p-value < 0.05, reject H\textsubscript{0}  (there is a significant change in intended behaviors). If p-value >= 0.05, fail to reject H\textsubscript{0} (no significant change).
    \end{enumerate}
    
    \begin{table}[H]
    \centering
    \resizebox{\columnwidth}{!}{%
    \begin{tabular}{|c|c|c|}
    \hline
    \textbf{Mean Increase in Number of Actions} & \textbf{Wilcoxon Statistic} & \textbf{P-value} \\
    \hline
    1.64 & 57.5 & 0.00068 \\
    \hline
    \end{tabular}%
    }
    \caption{Wilcoxon test results for intended actions related to food waste reduction.}
    \end{table}
    
    \begin{figure}[H]
        \centering
        \includegraphics[width=\columnwidth]{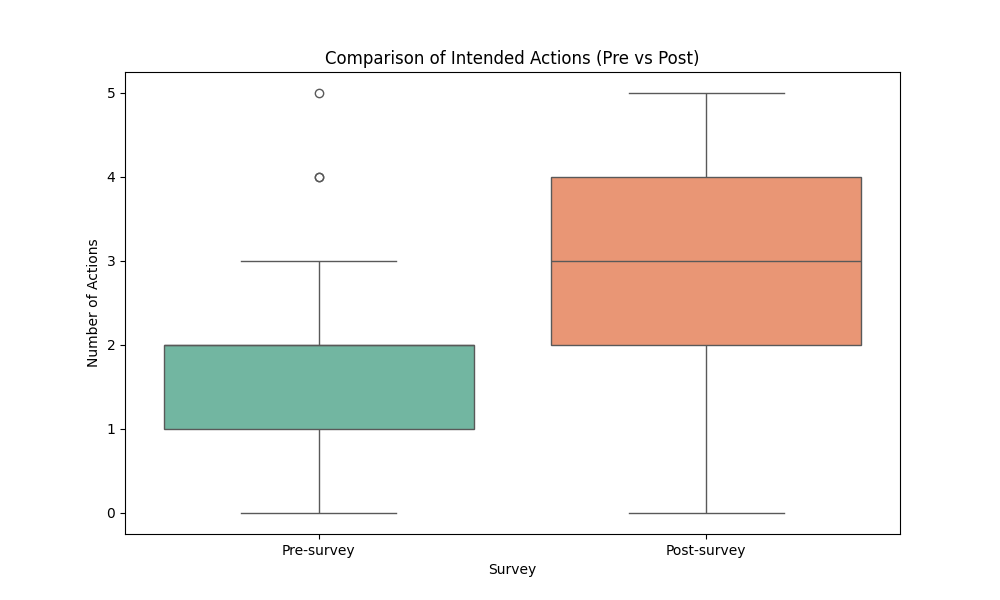}
        \caption{Distribution of Number of Reported Actions (Pre vs Post)}
    \end{figure}

    \item To assess the impact of the VR experience on participants' intended behaviors, we conducted a qualitative content analysis of open-ended survey responses collected before and after the intervention.
    \begin{enumerate}
    \item Participants answered the same open-ended question about food scrap disposal and reuse in both the pre- and post-survey. A total of 33 pre-survey and 30 post-survey valid responses were analyzed.
    \item Responses were manually reviewed and grouped into thematic categories. Pre-survey responses were categorized based on disposal methods (e.g., trash, composting), while post-survey responses were categorized by specific reuse ideas (e.g., composting, eggshell powder, cleaning agents, DIY uses).
    \item For each participant, we identified whether their post-survey response included reuse ideas not mentioned in their pre-survey response. If a participant mentioned any new, specific reuse application (e.g., “eggshell as calcium powder”), their response was counted as reflecting a behavioral shift.
    \item The number and percentage of participants showing a change were computed by comparing pre- and post-survey responses. Only clearly new and specific ideas were considered evidence of changed intent.
    \end{enumerate}
\end{enumerate}

\end{document}